\documentclass[conference]{IEEEtran}
\IEEEoverridecommandlockouts
\usepackage{cite}
\usepackage{amsmath,amssymb,amsfonts}
\usepackage{algorithmic}
\usepackage{graphicx}
\usepackage{textcomp}
\usepackage{xcolor}
\usepackage{subcaption}
\usepackage{enumitem}
\usepackage{multirow}
\usepackage{tabularx}
\usepackage{makecell}
\usepackage[pagebackref=false,breaklinks=true,colorlinks,bookmarks=false]{hyperref}
\def\BibTeX{{\rm B\kern-.05em{\sc i\kern-.025em b}\kern-.08em
    T\kern-.1667em\lower.7ex\hbox{E}\kern-.125emX}}
    
\begin{document}

\title{Smartphone-based Iris Recognition through High-
Quality Visible Spectrum Iris Capture}

\author{\IEEEauthorblockN{1\textsuperscript{st} Naveenkumar G Venkataswamy}
\IEEEauthorblockA{\textit{Department of Electrical and Computer Engineering} \\
\textit{Clarkson University}\\
New York, USA \\
venkatng@clarkson.edu}

\and
\IEEEauthorblockN{2\textsuperscript{nd} Yu Liu}
\IEEEauthorblockA{\textit{Department of Computer Science} \\
\textit{Clarkson University}\\
New York, USA \\
liuy5@clarkson.edu }

\and
\IEEEauthorblockN{3\textsuperscript{rd} Surendra Singh}
\IEEEauthorblockA{\textit{Department of Electrical and Computer Engineering} \\
\textit{Clarkson University}\\
New York, USA \\
sursing@clarkson.edu}

\and
\IEEEauthorblockN{4\textsuperscript{th} Soumyabrata Dey}
\IEEEauthorblockA{\textit{Department of Computer Science} \\
\textit{Clarkson University}\\
New York, USA \\
sdey@clarkson.edu }

\and
\IEEEauthorblockN{5\textsuperscript{th} Stephanie Schuckers}
\IEEEauthorblockA{\textit{Department of Electrical and Computer Engineering} \\
\textit{Clarkson University}\\
New York, USA \\
sschucke@clarkson.edu }

\and
\IEEEauthorblockN{6\textsuperscript{th} Masudul H Imtiaz}
\IEEEauthorblockA{\textit{Department of Electrical and Computer Engineering} \\
\textit{Clarkson University}\\
New York, USA \\
mimtiaz@clarkson.edu }
}

\IEEEoverridecommandlockouts
\maketitle
\IEEEpubidadjcol 

\begin{abstract}
Iris recognition is widely acknowledged for its exceptional accuracy in biometric authentication, traditionally relying on near-infrared (NIR) imaging. Recently, visible spectrum (VIS) imaging via accessible smartphone cameras has been explored for biometric capture. However, a thorough study of iris recognition using smartphone-captured 'High-Quality' VIS images and cross-spectral matching with previously enrolled NIR images has not been conducted. The primary challenge lies in capturing high-quality biometrics, a known limitation of smartphone cameras. This study introduces a novel Android application designed to consistently capture high-quality VIS iris images through automated focus and zoom adjustments. The application integrates a YOLOv3-tiny model for precise eye and iris detection and a lightweight Ghost-Attention U-Net (G-ATTU-Net) for segmentation, while adhering to ISO/IEC 29794-6 standards for image quality. The approach was validated using smartphone-captured VIS and NIR iris images from 47 subjects, achieving a True Acceptance Rate (TAR) of 96.57\% for VIS images and 97.95\% for NIR images, with consistent performance across various capture distances and iris colors. This robust solution is expected to significantly advance the field of iris biometrics, with important implications for enhancing smartphone security. 

\end{abstract}


\section{Introduction}

Over the past decade, the use of smartphones has become increasingly popular, alongside growing security concerns. Traditional authentication methods, such as passwords and personal identification numbers (PINs), have been found vulnerable to a variety of security attacks \cite{markert2020pin}. These vulnerabilities have raised the need for more robust and secure authentication methods. The integration of biometric technology into personal devices like smartphones holds significant promise for improving personal authentication, thereby improving security and user experience concerns. Fingerprint and face recognition technologies have revolutionized the way individuals unlock their smartphones and authorize transactions \cite{spolaor2016biometric}. Nevertheless, iris recognition represents an equally promising alternative within the realm of biometrics. Iris patterns are inherently distinctive and consistent, making them an excellent choice for personal identification purposes \cite{bowyer2008image, daugman2009iris}. Despite its remarkable performance, the widespread adoption of iris recognition technology in mobile devices, such as smartphones and tablets, remains relatively rare. Several factors contribute to this underutilization, including the dependence on near-infrared imaging of iris recognition, which operates within the 700 to 900 nm wavelength range, and the need for additional camera installation, as smartphones mostly feature VIS imaging cameras that operate within the 400 to 700 nm wavelength range ~\cite{raja2015smartphone}.

\begin{table*}[!ht]
\centering
\caption{Publicly available VIS iris datasets}
\label{table:datasets}
\resizebox{\textwidth}{!}{
\begin{tabular}{|c|c|l|l|c|c|c|c|}
\hline
\textbf{Name} & \textbf{Camera Type} &\textbf{Sensor} &\textbf{Environment}& \textbf{Number of Subjects}& \textbf{Images Total} &\textbf{Custom Application} &\textbf{ISO Quality checks} \\
\hline
CSIP & VIS & \makecell{Sony Ericsson Xperia Arc S\\ Apple iPhone 4\\ THL W200\\ Huawei Ideos X3 (U8510)} & Unconstrained & 50 & 2004 & No & No \\ 
\hline
MICHE - I/II & VIS & \makecell{iPhone 5\\ Samsung Galaxy IV\\ Galaxy Tablet II} & Unconstrained & 92 & 3732 & No & No \\ 
\hline
VSIOB & VIS & \makecell{iPhone 5s\\ Samsung Note 4\\ Oppo N1} & Indoor and Unconstrained & 550 & NA& No & No \\ 
\hline
IIITD CMPD & VIS & MicroMax A350 Canvas & Indoor & 145 & 2380 & No & No \\
\hline
IrisCorneaDataset & VIS & USB Webcam & Indoor & 39 & 780 & No & No \\
\hline
UBIRIS-V1 & VIS & Nikon E5700 & Indoor & 241 & 1877 & No & No \\
\hline
UBIRIS-V2 & VIS & Canon EOS 5D & Unconstrained & 261 & 11102 & No & No \\\hline
CUVIRIS & VIS, NIR& Samsung Galaxy S21 Ultra, IG-AD100 & Indoor & 47 & 940 & Yes & Yes \\\hline
\end{tabular}}
\vspace{-3mm}
\end{table*}

One of the main reasons NIR imaging is favored in traditional iris recognition systems is because of the way different types of light interact with the iris. The human iris absorbs and reflects light differently depending on its wavelength. In VIS (400-700 nm), the iris, especially darker irises, tends to absorb more light, which lowers the contrast between the iris and the pupil and creates stronger reflections. This can make it harder to capture clear and detailed iris patterns, particularly in uncontrolled lighting environments. Conversely, in the near-infrared spectrum (700–900 nm), the iris absorbs less light and reflects more, enhancing the contrast between the iris and other eye structures. This improved contrast leads to clearer, more detailed iris images, facilitating accurate recognition across diverse populations \cite{hosseini2010pigment}.

Specialized NIR imaging equipment is typically used to capture iris images because NIR illumination enhances iris texture, reduces reflections, and mitigates environmental effects such as ambient light \cite{winston2019comprehensive}. However, visible-light imaging is less prevalent due to its susceptibility to reflections and environmental artifacts. In uncontrolled environments, or when the subject is uncooperative, iris images captured in the VIS often suffer from issues like optical and motion blurs, reflections, and other artifacts, which impair recognition performance. Modern smartphones come with high-resolution cameras and smart image processing features that help overcome these challenges. With features such as adjustable focus, optical image stabilization, and powerful processors, these devices can capture high-quality iris images, even under varying conditions. By making the most of these advancements, we can create reliable iris recognition systems that are both accessible and user-friendly.

Previous research on VIS-based iris recognition (Section II) utilized low-quality images captured by various old smartphone camera models under constrained and unconstrained conditions, including low light, shadows, off-angle gaze directions, optical and motion blur, and out-of-focus issues ~\cite{rattani2016icip, de2015mobile, santos2015fusing, edwards2012quantitative , proencca2009ubiris}. Although multiple smartphone-based VIS iris image databases exist, these collections were obtained using old smartphones and low-quality cameras and default inbuilt camera applications without dedicated application and quality control measures, resulting in poor-quality images. Modern day's smartphones have higher-resolution cameras and higher computational capabilities. We can take advantage of the latest smartphones for the iris recognition process.

In this study, we develop a dedicated Android smartphone application that leverages the software and hardware capabilities of the latest smartphone devices to capture high-quality VIS iris images. The key features of the application are:

1. Eye and Iris Detection: The application utilizes an eye detection model based on the YOLOV3-tiny architecture to detect the location of the eye and the iris in real time (see Figure~\ref{fig:Add an image of yolov3 eye detection}).
   
\begin{figure}[h]
\centering
    \includegraphics[width=9cm]{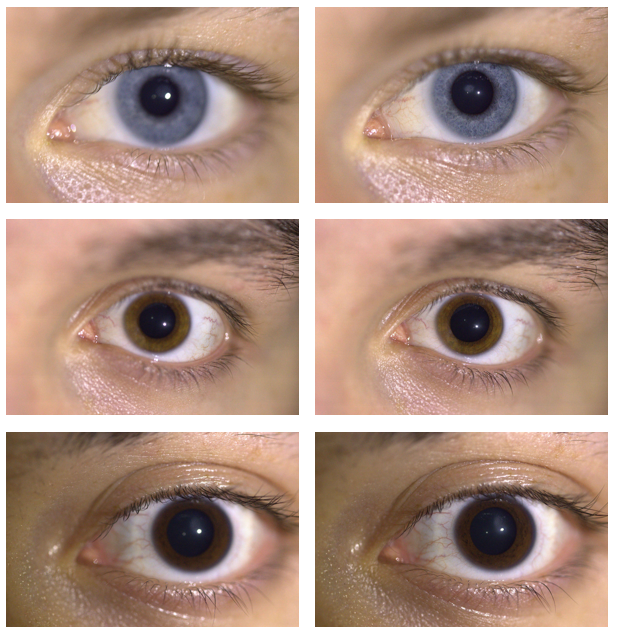}
    \caption{\footnotesize Automatic Focal Point Adjustment. The left column shows out-of-focus eye images before the focal point adjustment, while the right column shows the corresponding images after the focus is automatically adjusted to the iris region.}
    \label{fig:focusset}
    \vspace{-3mm}
 \end{figure}

2. Automatic Focal Point Adjustment: Once the iris is detected, the focal point of the camera is automatically adjusted to focus on the iris region (Figure~\ref{fig:focusset}).

3. Automatic Zooming: The application automatically controls the zoom level based on the size of the detected eye’s bounding box, ensuring consistent image capture regardless of the distance between the subject and the camera.

4. Consistent eye framing and focus: A feedback loop adjusts both the zoom (using eye coordinates) and the focus (using iris coordinates) to ensure that the eye remains consistently well framed and sharply focused.

5. Eye Image Cropping: After focusing and zooming, the eye image is cropped to a consistent size of 640x480 pixels, ensuring uniformity in image dimensions for further processing.

6. Image Quality Feedback: The application provides real-time feedback to the user if the captured image does not meet the required quality standards.

7. Image Quality Checks: The application incorporates image quality checks according to the ISO/IEC 29794-6 standards \cite{iris_standard_report}, ensuring that only high-quality images are retained.

8. Iris Segmentation: A custom-developed G-ATTU-Net model is integrated into the application for real-time iris segmentation on the device.

To evaluate the performance of the application, we collected iris images from 47 participants (Section IV) using a Samsung Galaxy S21 Ultra smartphone with a built-in flashlight as the default illumination source. While the dataset may seem relatively small, it is the first publicly available dataset that focuses on high-resolution VIS iris images captured using modern smartphones with built-in quality checks and real-time segmentation. This dataset, will be made available to the research community for non-commercial use, enabling future research into VIS iris recognition and cross-spectral matching. Additionally, the technical contributions of this work lie in the development of an efficient mobile app that automatically adjusts zoom and focus to ensure consistent iris image capture, regardless of subject distance. The G-ATTU-Net model provides a lightweight and efficient solution for segmenting the iris on mobile devices, enabling real-time processing. This work contributes significantly to advancing mobile-based biometric systems and offers valuable resources for improving VIS iris recognition methods.

The rest of the paper is organized as follows: Section II discusses related work on VIS iris recognition, including the limitations of existing approaches. Section III details the models employed in our proposed system, including the YOLOv3-tiny and G-ATTU-Net architectures. Section IV presents the experimental setup, data collection methodology, and the results obtained from evaluating our system on the collected dataset. Section V provides an analysis and discussion of the results, and Section VI concludes the paper with final remarks and potential directions for future research.

\section{Related work}
Several VIS smartphone-based iris datasets have been developed for iris recognition research. The VISOB dataset \cite{rattani2016icip} was collected using the front and back cameras of the Samsung Note 4, iPhone 5S, and Oppo N1, with a capture distance of approximately 8 to 12 inches under three different lighting conditions: regular office light, dim light, and natural daylight.

The MICHE I and II datasets \cite{de2015mobile} were collected using the iPhone 5, Samsung Galaxy IV, and Galaxy Tablet II in unconstrained environments. Similarly, the CSIP dataset \cite{santos2015fusing} was created using four different smartphones: the Xperia Arc S, iPhone 4, THL W200, and Huawei Ideos X3, all under unconstrained conditions. The IITD CMPD was collected under indoor conditions using the MicroMax A350 Canvas smartphone. Additionally, VIS cameras were used to capture iris images for the UBIRIS-V1 \cite{edwards2012quantitative} dataset, which was collected using the Nikon E5700 camera, and the UBIRIS-V2 dataset \cite{proencca2009ubiris}, which was collected using the Canon EOS 5D camera. The IrisCorneaDataset was collected using a USB webcam in indoor conditions.

All these datasets were collected using older smartphone cameras, without dedicated applications, and often required the subject's cooperation. In contrast, our Clarkson University Visible Iris (CUVIRIS) dataset was collected using the latest Samsung Galaxy S21 Ultra smartphone, with a dedicated application that includes rigorous quality checks. The corresponding NIR images were also captured using the IrisGuard IG-AD100 \cite{IrisGuard} in indoor conditions with a capture distance of up to 50 cm, using the mobile flashlight as an illumination source. Table~\ref{table:datasets} provides detailed information about the publicly available VIS iris datasets, including the camera sensors used, the number of images and participants, the use of customized applications, and the quality checks performed.

Recent studies have demonstrated significant advances in VIS-based iris recognition. One study \cite{raja2015smartphone} utilized two smartphone models, the iPhone 5S and Nokia Lumia 1020, to create the VSSIRIS database, which was captured under mixed illumination conditions. The proposed segmentation technique, which employs saliency maps and anisotropic diffusion to accurately localize and segment the iris region, achieved up to 85\% accuracy with the standard OSIRIS v4.1 toolkit and yielded an Equal Error Rate (EER) of 1.62\%. Although this method demonstrated significant performance, its EER in real-world scenarios indicates that further improvements are needed.

Another study \cite{raja2015iris} used a white LED as an illumination source to create a database of NIR and VIS iris images using smartphones such as the iPhone 5S, Nokia Lumia 1020, and Samsung Active S4. This study reported a Genuine Match Rate (GMR) of 91.01\% at a False Match Rate (FMR) of 0.01\%. However, the external LED light source focused illumination on certain parts of the iris, which enhanced texture visibility but led to a loss of information where the light incident on the iris region.

The work \cite{proenca2009iris} proposed a segmentation method that showed error rates ranging from 1.87\% to 5.02\% on the UBIRIS-V2 dataset \cite{proencca2009ubiris}. Despite its effectiveness, the method’s accuracy was compromised in cases of severe occlusion, extreme off-angle images, and poor lighting conditions. Additionally, the performance of this method is highly dependent on the quality and type of input data, resulting in varying results across different datasets.

Another study \cite{trokielewicz2016iris} collected a new dataset using the iPhone 5S and evaluated the performance using state-of-the-art recognition algorithms. The study achieved a Correct Match Rate (CMR) of over 99.5\% and a zero False Match Rate (FMR) across all methods, with a Failure to Enroll (FTE) rate of 1.27\%. The authors noted that incorrect segmentation contributed significantly to recognition errors, highlighting the critical impact of iris sample quality on segmentation and overall recognition performance.

Despite these advances, the need for higher-quality VIS iris datasets captured under real-world conditions remains. This study addresses this gap by introducing the CUVIRIS dataset and proposing advanced techniques for iris image capture and segmentation. Specifically, we developed a lightweight Ghost-Attention U-Net (G-ATTU-Net) model designed to perform accurate iris segmentation on resource-constrained mobile devices, such as smartphones. The model is optimized for efficiency, reducing the number of parameters while maintaining high segmentation accuracy, making it suitable for real-time deployment on mobile platforms.

\begin{figure*}[h]
\centering
    \includegraphics[width=1\textwidth]{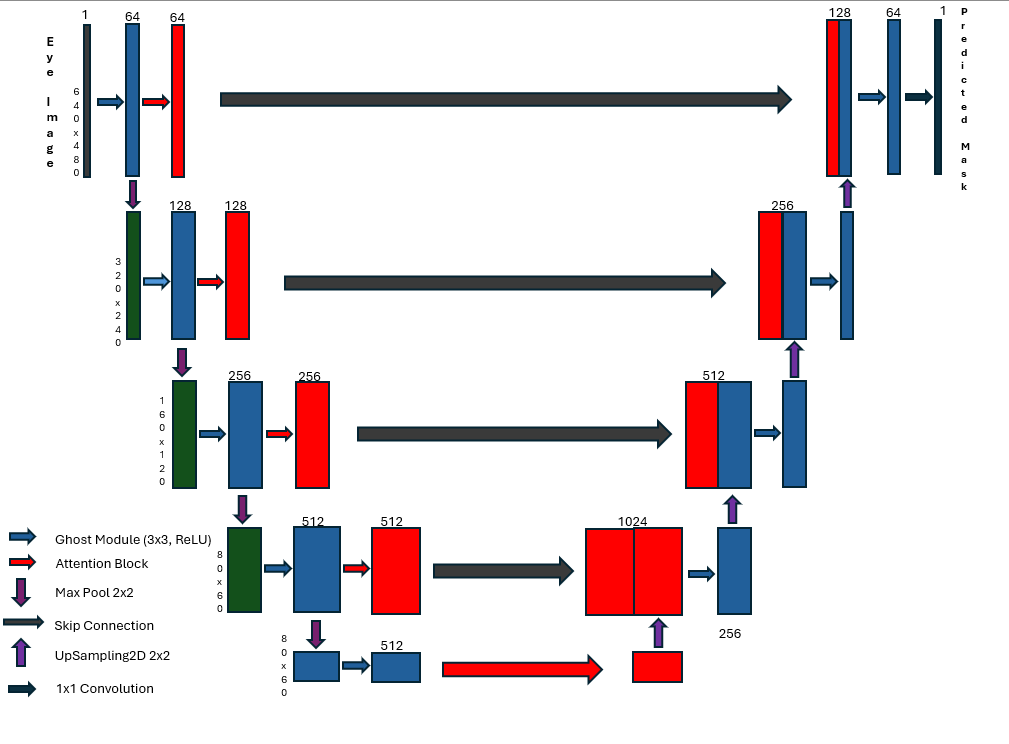}
    \caption{\footnotesize The architecture of the proposed G-ATTU-Net model. Each dark blue block represents a multi-channel feature map produced by a Ghost Module, with the numbers above indicating the number of channels and the vertical numbers on the left indicating the spatial dimensions (height x width). The red blocks indicate Attention Blocks used to refine features before concatenation in the decoder path.}
    \label{fig:UnetArch}
    \vspace{-3mm}
 \end{figure*}

\section{Model Development}
\subsection{YOLOV3-Tiny Eye and Iris Detection Model}
Considering the need to integrate this model into a smartphone while maintaining the performance of the Android application, we opted for a compact object detection model, specifically the YOLOv3-Tiny \cite{adarsh2020yolo}. This model was chosen due to its lower parameter count, faster detection speed, and overall efficiency, featuring 24 layers with 0.557 GFLOPS and capable of processing at 220 FPS.

The training was carried out using the transfer learning technique with more than 3,000 eye images from the publicly available UBIRIS-V1 \cite{edwards2012quantitative} and UBIRIS-V2 datasets \cite{proencca2009ubiris} and over 750 images from our own smartphone-captured dataset, divided into an 80\% training set and a 20\% testing set by resizing the images into 416x416 as model input. Fine-tuning of the model was done using the Nvidia GeForce RTX 3080, achieving a training accuracy of over 98\% and a testing accuracy of over 97\%. The resulting weights and configuration were saved in Hierarchical Data Format (HDF5) \cite{folk2011overview}.

Given the resource-constrained environment of a smartphone, we performed quantization to optimize the model for deployment. Quantization involves reducing the precision of model parameters and activations from high-precision floating point (typically FP32) to lower-precision formats such as 8-bit integer (int8) format, which significantly reduces the computational load and memory footprint. Specifically, we quantized the previously trained model from the HDF5 format to TensorFlow Lite (TFLite) format, a process that involved mapping floating-point values to discrete integer levels defined by quantization step size and zero-point parameters. This conversion allows for efficient execution on mobile devices while preserving model accuracy.

A comparison between the original HDF5 model and the quantized TFLite model revealed that the TFLite model reduced the file size from 54 MB to 33 MB and improved detection time on a standard GPU, in this case, a Nvidia RTX 3080, from 0.1 to 0.06 ms. For deployment, we used a Samsung Galaxy S21 smartphone, which comes with a built-in Adreno 660 GPU, a high-performance GPU known for its efficient graphics rendering and AI processing capabilities. On this device, the detection time was further reduced from 0.26 ms to 0.12 ms, demonstrating the substantial performance benefits achieved through quantization and confirming the suitability of the model for real-time eye and iris detection on resource-constrained mobile platforms.

\subsection{G-ATTU-Net Segmentation Model}
Iris recognition systems typically involve five basic steps: iris image acquisition, preprocessing, segmentation, feature extraction, and matching. Accurate segmentation of the iris is crucial as it enables the extraction of valuable information from the iris image, thereby improving the overall accuracy of the iris recognition system. Traditional segmentation methods, such as the Hough Transform (HT) and Daugman's integro-differential operator, primarily rely on detecting circular boundaries in the eye image. These methods first locate the pupil as the inner boundary of the iris and then identify other parameters, such as the eyelid and limbic regions, to separate them from the iris. While these methods are effective for near-infrared (NIR) images where the contrast between the iris and other parts of the eye is high, they struggle in visible light environments. Factors like environmental noise, occlusions, reflections, and non-circular boundaries limit the accuracy of traditional methods in visible light images, necessitating the development of more advanced techniques, such as deep learning-based approaches.

\begin{figure}[h]
\centering
    \includegraphics[width=9cm]{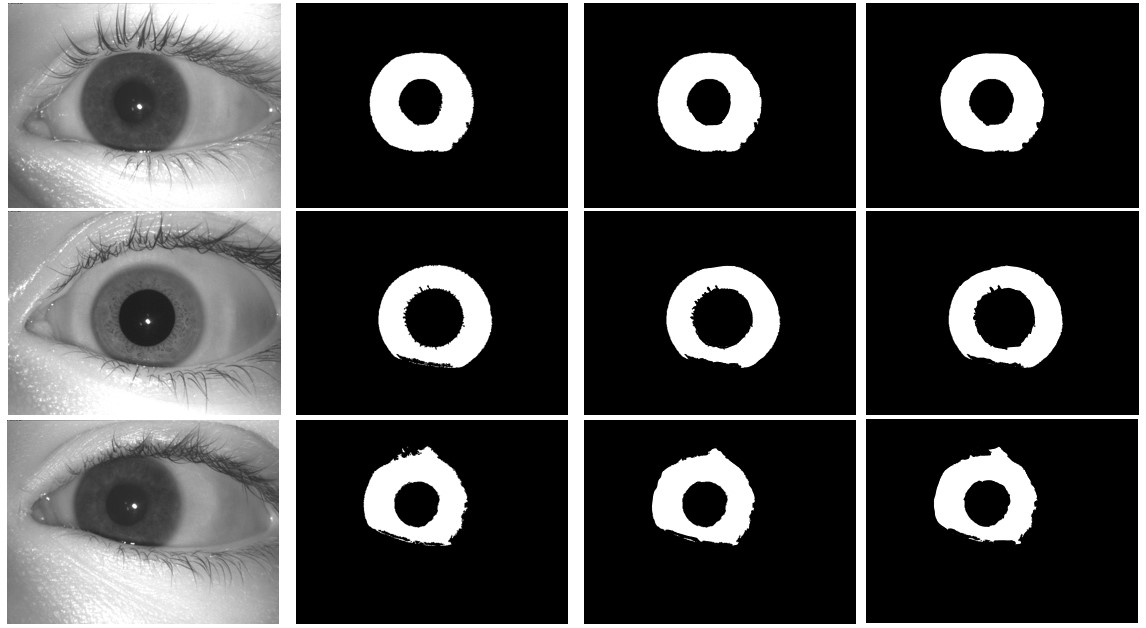}
    \caption{\footnotesize The input images and segmentation results using both the standard Attention U-Net and the GAtt-UNet. The first column shows the red channel images extracted from RGB inputs after cropping the eye region using the YOLOv3-Tiny eye and iris detection model from the CUVIRIS dataset. The second column contains the ground truth masks for the iris region. The third column displays the segmentation results from the standard ATT-UNet, and the fourth column shows the results from the G-ATTU-Net.}
    \label{fig:Unetmasks}
    \vspace{-3mm}
 \end{figure}
 
Several works have explored deep learning-based networks for iris segmentation. For instance, the study by \cite{arsalan2017deep} introduces a two-stage iris segmentation method that combines a modified Circular HT for initial rough boundary detection with a Convolutional Neural Network (CNN) for segmentation within the region of interest (ROI). This approach effectively handles noisy, non-ideal visible light images and outperforms traditional methods, achieving lower segmentation errors and higher accuracy on datasets such as NICE-II. However, the method's effectiveness can be limited by severe occlusions and extreme off-angle images, leading to segmentation inaccuracies. Another study \cite{jan2024iris} proposes an iris segmentation method that integrates the Viola-Jones algorithm with geometric information of the human face to accurately segment eyes, followed by a combination of HT and Lagrange interpolation polynomial for iris contour localization. This method demonstrated high accuracy, achieving up to 99.3\% accuracy in localizing the iris outer contour on the CASIA-IrisV4-Distance dataset and 98.51\% on the IITD-V1.0 dataset. However, the method faces limitations in dealing with images containing eyeglasses, partial faces, or non-frontal face images, and further refinement is needed to handle reflections that interfere with boundary regularization.

Many U-Net architecture-based segmentation models have also been introduced in the current research; however, they often face challenges such as low accuracy, slow runtime, and high computational complexity, which are not feasible for resource-constrained devices like smartphones. The proposed method in \cite{zhang2019robust} uses an improved U-Net architecture with dilated convolution for iris segmentation, comprising a U-shaped network with multiple convolutional and pooling layers. The FD-UNet model, consisting of 13 convolutional layers and 5 pooling layers with dilated convolutions, increases the receptive field without significantly adding to the number of parameters. The model achieved an F1 score of 97.36\% on the CASIA-iris-interval-v4.0 dataset but is relatively large in terms of computational requirements due to the complexity of the architecture. Another study \cite{lian2018attention} proposes an Attention U-Net (ATT-UNet), which builds upon the original U-Net architecture by incorporating an attention mechanism. The ATT-UNet includes a contracting path based on the VGG16 architecture (without fully connected layers) and an expanding path for segmentation. This model, consisting of 13 convolutional layers and approximately 31 million network parameters, uses a bounding box regression module to estimate the iris region, guiding the attention mechanism. Although the attention-guided U-Net achieves high segmentation accuracy, its large model size makes it computationally intensive.

In this work, we present a lightweight attention-based U-Net specifically designed for accurate iris segmentation on resource-constrained mobile devices. Our proposed architecture, as shown in Fig.~\ref{fig:UnetArch}, referred to as Ghost-Attention U-Net (G-ATTU-Net), is optimized to be approximately a quarter of the size of the traditional ATT-UNet, while maintaining comparable performance in segmentation accuracy. The G-ATTU-Net architecture comprises several key components that contribute to its efficiency and performance, with over 8 million parameters. The network accepts a single-channel grayscale image of size 640x480 as input, corresponding to the iris image captured by mobile devices. The encoder path of the network includes four stages, each containing Ghost Modules followed by a Max Pooling layer. Ghost Modules generate efficient feature maps by combining a standard convolution (Conv2D) with a depthwise separable convolution (DepthwiseConv2D). This setup, along with batch normalization and ReLU activation, enables the network to achieve richer feature representations with fewer parameters. The number of filters in the encoder path starts at 64 and doubles at each stage, progressing through 128, 256, and 512 filters, with Max Pooling layers reducing the spatial dimensions by half at each stage.

At the network's deepest point, the bottleneck, a Ghost Module with 512 filters, is used to extract the most abstract features without further downsampling. The decoder path mirrors the encoder, utilizing upsampling layers to gradually restore the spatial dimensions. Each stage in the decoder includes a Ghost Module, with the number of filters halving as the spatial dimensions increase, moving from 512 back down to 64. Importantly, skip connections from the corresponding encoder layers are integrated into the decoder via Attention Blocks. These Attention Blocks, incorporated into the skip connections, play a crucial role in refining the features passed from the encoder to the decoder. They modulate the importance of features using a combination of 1x1 convolutions, ReLU activations, and sigmoid gating mechanisms, allowing the network to focus on the most relevant features and improve segmentation accuracy without significantly increasing model complexity. The final layer of the network is a 1x1 convolutional layer that outputs a single-channel binary mask of the segmented iris region, matching the input size of 640x480.

The G-ATTU-Net achieves a significant reduction in the number of parameters by leveraging Ghost Modules, which efficiently generate feature maps with fewer operations. The use of Attention Blocks ensures that the model remains focused on relevant features, compensating for the reduced parameter size while maintaining high segmentation accuracy. As a result, the G-ATTU-Net is particularly well-suited for real-time iris segmentation tasks on mobile and embedded devices, offering a powerful yet efficient solution for applications requiring both accuracy and computational efficiency.

\subsubsection{Training and Evaluation}
Both the standard Attention U-Net and the proposed G-ATTU-Net were initially trained on the CSIP, MICHE I and II , UBIRIS-V1 and V2, VSIOB, IrisCorneaDataset datasets, which together comprise over 20,000 images.

The eye region of each image was first cropped using the YOLOv3-Tiny eye and iris detection model, and the red channel was extracted from the RGB images to serve as input for the segmentation models. The dataset was split into 80\% for training and 20\% for validation. To enhance the robustness of both models, various data augmentation techniques were applied, including rotation, width and height shifts, and zooming. After the initial training phase, both models were fine-tuned on the CUVIRIS dataset. The training process employed the Adam optimizer with a learning rate of 0.0001. The Binary Cross entropy loss function was used, with accuracy as the primary metric for model evaluation.

The models' performance was evaluated on the CUVIRIS dataset in terms of mean Intersection over Union (IoU) and F1 score. The standard Attention U-Net achieved a mean IoU of 98.92\% and an F1 score of 98.56, indicating high accuracy in segmentation. In comparison, the G-ATTU-Net achieved a mean IoU of 97.16\% and an F1 score of 97.25, demonstrating slightly lower but still competitive performance with a significantly reduced parameter size. Fig.~\ref{fig:Unetmasks} shows the masks generated by both models.

\section{Application Development}
We utilize Android Studio IDE, selecting Java as the primary programming language for developing the Android application. The application was designed to target Android version 5.0 (Lollipop) and API level 21 and above, ensuring compatibility with a wide range of Android devices. 
\subsection{Application Components}
\subsubsection{User Interface}
To capture participant details, we designed a User Interface (UI) where users can input information such as Subject ID, the choice of left or right eye, a selection option for spoofing with an additional field to specify the type of spoof if applicable, and Session ID and Trial Number to differentiate iris images from the same participant. The collected subject information is saved for later use in naming the images according to the format 'subjectID-left/right-spooftype-sessionID-trialNumber.jpg'. Fig.~\ref{fig:UI_Preview}(a) showcases the developed interface.

\subsubsection{Image Quality Assessment}
We use the OpenCV library in two distinct ways to assess the basic quality of iris images. The first method, \textit{the} \textit{Laplacian}, examines variations in pixel intensity across the image to determine how blurry the picture is. This method calculates the image's second derivative to evaluate its sharpness. The second method, the \textit{Fast Fourier Transform (FFT)}, is also integrated into OpenCV to assess image sharpness. FFT converts an image from the spatial domain to the frequency domain, allowing for the examination of high-frequency components, which provides information about image sharpness. By combining these methods within the OpenCV library, the sharpness and blurriness of iris images are comprehensively evaluated.

\subsubsection{ISO/IEC 29794-6 Standards}
In the Android application, we have integrated the ISO/IEC 29794-6:2015 standards for iris image quality. This integration includes the implementation of key metrics such as Overall Quality, Grayscale Utilization, Iris-Pupil Contrast, Iris-Pupil Concentricity, Iris-Pupil Ratio, Iris-Sclera Contrast, Margin Adequacy, and Pupil Boundary Circularity. The detailed explanations of each metric are provided below:

\begin{enumerate}
    \item \textbf{Overall Quality}: Ranging from 0 to 100, this score is based on the product of normalized individual ISO metrics.
    \item \textbf{Grayscale Utilization}: Ranging from 0 to 20, it indicates the spread of intensity values within the iris portion of the image.
    \item \textbf{Iris-Pupil Contrast}: A value ranging from 0 to 100, representing image characteristics at the boundary between the iris and pupil region.
    \item \textbf{Iris-Pupil Concentricity}: A value ranging from 0 to 100, indicating the degree to which the pupil center and iris center coincide.
    \item \textbf{Iris-Pupil Ratio}: Values ranging from 9.58 to 121.30, depicting the degree of pupil dilation or constriction.
    \item \textbf{Iris-Sclera contrast}: A value ranging from 0 to 100, representing image characteristics at the boundary between the iris region and the sclera.
    \item \textbf{Margin Adequacy}: A value ranging from 0 to 100, indicating the degree to which the iris portion is centered relative to the image edges.
    \item \textbf{Pupil Boundary Circularity}: This is a value ranging from 0 to 100 that represents the circularity of the iris-pupil boundary.
    \item \textbf{Sharpness}: A value ranging from 0 to 100, indicating the degree of focus in the image.
    \item \textbf{Usable Iris Area}: A value ranging from 0 to 100, representing the fraction of the iris portion not occluded by eyelids, eyelashes, or specular reflections.
\end{enumerate}

Table~\ref{table:quality metrics} summarizes the metrics used, their ranges, the recommended values, and the thresholds we considered. These metrics are employed to assess and ensure the quality of the iris images captured by the application, in alignment with established standards.

\begin{table}[!ht]
\scriptsize
\scriptsize
\centering
\caption{{Quality metrics comparison as per ISO/IEC 29794-6 standards showing the recommended value, range and used threshold values for various metrics}}
\label{table:quality metrics}
\begin{tabular}{|c|c|c|c|}
\hline
 \textbf{Metric} &  \textbf{Range} &  \textbf{Recommended Value} & \textbf{Value threshold} \\
  \hline
 Overall Quality & 0-100 & Higher the better& \textgreater70\\ 
  \hline
 Grayscale Utilization & 0-20 & \textgreater6 & \textgreater6\\ 
  \hline
 Iris-Pupil Concentricity & 0-100 & \textgreater90 & \textgreater90\\ 
  \hline
 Iris-Pupil Contrast & 0-100 & \textgreater30 & \textgreater30\\
 \hline
 Iris-Pupil Ratio & 0-100 & 20-70 & \textgreater20\\
 \hline
 Iris-Sclera Contrast & 0-100 & \textgreater5 & \textgreater5\\
 \hline
 Margin Adequacy & 0-100 & \textgreater80 & \textgreater80\\
 \hline
 Pupil-Boundary Circularity & 0-100 & Higher the better & \textgreater70\\
 \hline
 Sharpness & 0-100 & Higher the better & \textgreater80\\
 \hline
 Usable Iris Area & 0-100 & \textgreater70 & \textgreater70\\
 \hline
\end{tabular}
\vspace{-3mm}
\end{table}

\begin{figure*}[h]
\centering
    \includegraphics[width=1\textwidth]{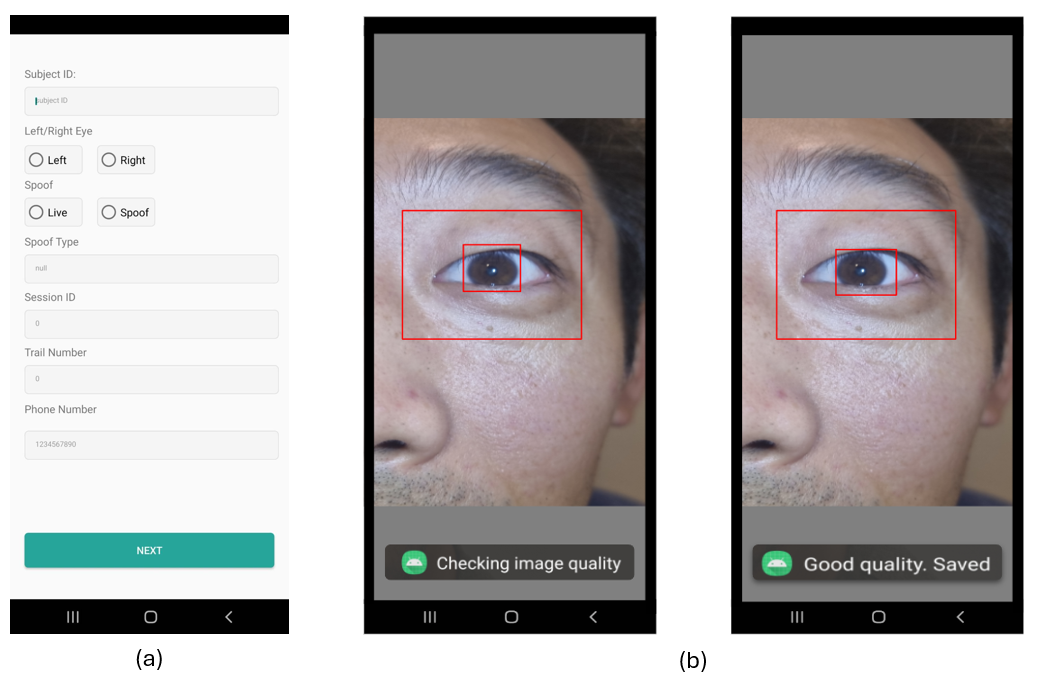}
    \caption{\footnotesize (a) User interface for inputting participant details and (b) camera preview displaying detected eye and iris bounding boxes with real-time user feedback, ensuring accurate and focused image capture.}
    \label{fig:UI_Preview}
    \vspace{-3mm}
 \end{figure*}
 
\subsubsection{YOLOV3-Tiny Eye and Iris Detection Model}
Before diving into Android application development, our primary focus was on creating a deep learning-based eye and iris detection model. Considering the need to integrate this model into a smartphone while maintaining the performance of the Android application, we opted for a compact object detection model, specifically the YOLOv3-Tiny \cite{adarsh2020yolo}. This model was chosen due to its lower parameter count, faster detection speed, and overall efficiency, featuring 24 layers with 0.557 GFLOPS and capable of processing at 220 FPS.

The training was carried out using the transfer learning technique with more than 3,000 eye images from the publicly available UBIRIS-V1 \cite{edwards2012quantitative} and UBIRIS-V2 datasets \cite{proencca2009ubiris} and over 750 images from our own smartphone-captured dataset, divided into an 80\% training set and a 20\% testing set by resizing the images into 416x416 as model input. Fine-tuning of the model was done using the Nvidia GeForce RTX 3080, achieving a training accuracy of over 98\% and a testing accuracy of over 97\%. The resulting weights and configuration were saved in Hierarchical Data Format (HDF5) \cite{folk2011overview}.

Given the resource-constrained environment of a smartphone, we performed quantization to optimize the model for deployment. Quantization involves reducing the precision of model parameters and activations from high-precision floating point (typically FP32) to lower-precision formats such as 8-bit integer (int8) format, which significantly reduces the computational load and memory footprint. Specifically, we quantized the previously trained model from the HDF5 format to TensorFlow Lite (TFLite) format, a process that involved mapping floating-point values to discrete integer levels defined by quantization step size and zero-point parameters. This conversion allows for efficient execution on mobile devices while preserving model accuracy.

A comparison between the original HDF5 model and the quantized TFLite model revealed that the TFLite model reduced the file size from 54 MB to 33 MB and improved detection time on a standard GPU—in this case, an Nvidia RTX 3080—from 0.1 ms to 0.06 ms. For deployment, we used a Samsung Galaxy S21 smartphone, which comes with a built-in Adreno 660 GPU, a high-performance GPU known for its efficient graphics rendering and AI processing capabilities. On this device, the detection time was further reduced from 0.26 ms to 0.12 ms, demonstrating the substantial performance benefits achieved through quantization and confirming the suitability of the model for real-time eye and iris detection on resource-constrained mobile platforms.

\subsubsection{G-ATTU-Net Segmentation Model}
Iris recognition systems typically involve five basic steps: iris image acquisition, preprocessing, segmentation, feature extraction, and matching. Accurate segmentation of the iris is crucial as it enables the extraction of valuable information from the iris image, thereby improving the overall accuracy of the iris recognition system. Traditional segmentation methods, such as the Hough Transform (HT) and Daugman's integro-differential operator, primarily rely on detecting circular boundaries in the eye image. These methods first locate the pupil as the inner boundary of the iris and then identify other parameters, such as the eyelid and limbic regions, to separate them from the iris. While these methods are effective for near-infrared (NIR) images where the contrast between the iris and other parts of the eye is high, they struggle in visible light environments. Factors like environmental noise, occlusions, reflections, and non-circular boundaries limit the accuracy of traditional methods in visible light images, necessitating the development of more advanced techniques, such as deep learning-based approaches.

\subsubsection{Application Flow}
\begin{figure}[h]
\centering
    \includegraphics[width=9cm, height=13cm]{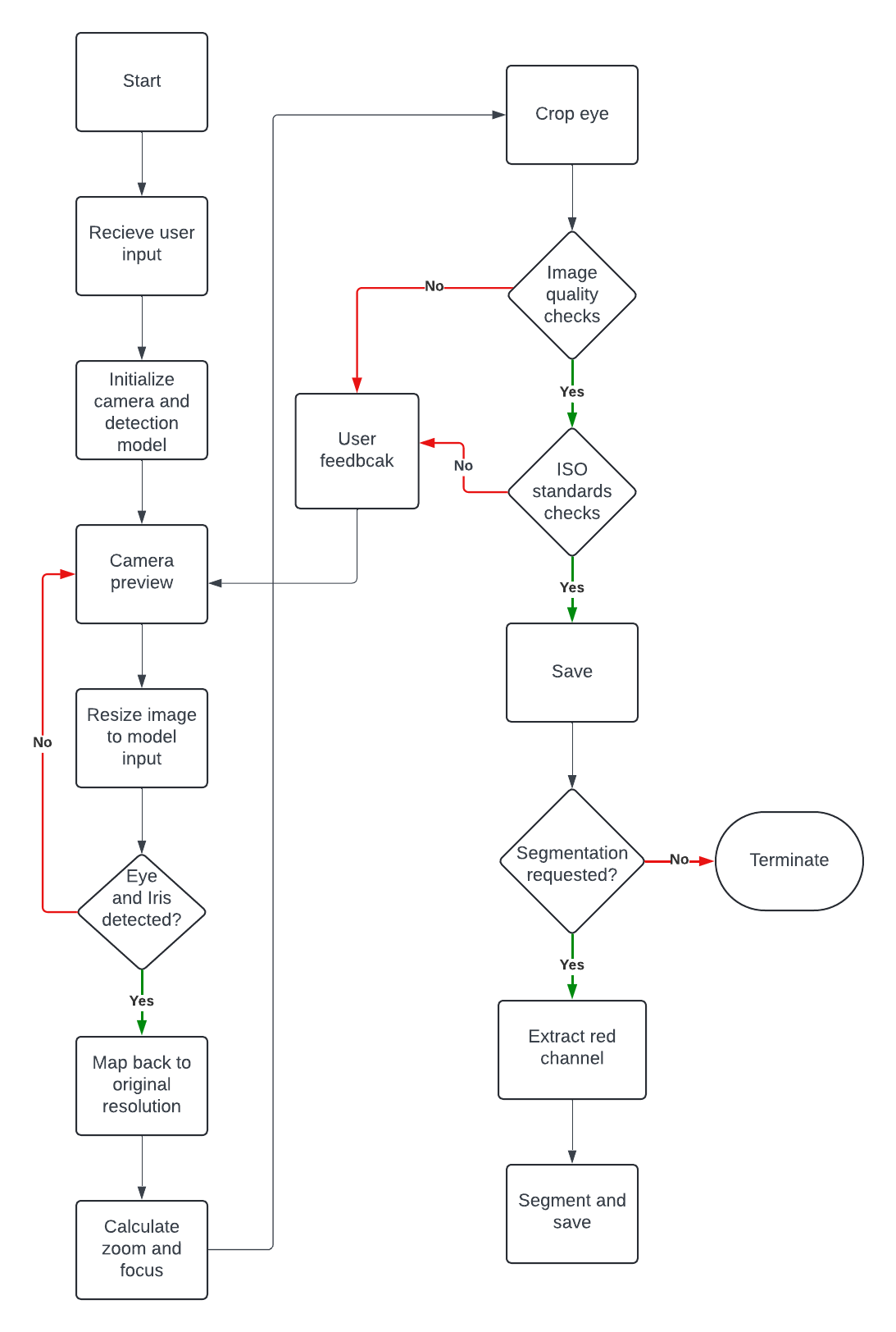}
    \caption{\footnotesize Process flow of the Android application for capturing and processing iris images.}
    \label{fig:appflow}
    \vspace{-3mm}
 \end{figure}

The overall process flow of the Android application is depicted in Fig.~\ref{fig:appflow}. When a user launches the application, the UI opens and asks the participant or the collector to enter the participant details as mentioned in Section IV.A1. In case any information is missing, the application requests the user to fill in the required details. After entering all the participant details, the camera permissions were requested, and after the permission was provided, the YOLOV3 eye and iris detection model (Section III.A) was initialized, and the camera preview was shown. Upon opening the camera preview, each frame from the camera is resized into 416x416 and passed as input to the eye and iris detection model. If there is no eye and iris detected on the frame, the process loops until the eye and iris are detected. The model provides two sets of coordinates: one for the eye and another for the iris. These coordinates are crucial for dynamically adjusting both the camera’s zoom level and focal length. To accurately map these coordinates from the resized image back to the original camera sensor image dimensions, we implemented a process that utilizes scaling factors based on the ratio of the original image size to the resized image size. This ensures that both the eye and iris key points identified by the model in the resized image correspond correctly to their positions in the original high-resolution image. Given the original image dimensions $W_{\text{original}}$ and $H_{\text{original}}$, and the resized image dimensions $W_{\text{resized}}$ and $H_{\text{resized}}$, the scaling factors are calculated as:

\[
\text{Scale\_factor\_x} = \frac{W_{\text{original}}}{W_{\text{resized}}}
\]
\[
\text{Scale\_factor\_y} = \frac{H_{\text{original}}}{H_{\text{resized}}}
\]

For any detected coordinates $(x, y)$ in the resized image, the corresponding coordinates in the original image are then computed using the following equations:

\[
x_{\text{original}} = x \times \text{Scale\_factor\_x}
\]
\[
y_{\text{original}} = y \times \text{Scale\_factor\_y}
\]

The eye coordinates are used to determine the bounding box around the eye on the original high-resolution image. The bounding box dimensions are calculated as:

\[
\text{width}_{\text{bbox}} = x_{\text{max}} - x_{\text{min}}
\]
\[
\text{height}_{\text{bbox}} = y_{\text{max}} - y_{\text{min}}
\]

A target bounding box size is defined to represent the ideal eye and iris size in the image frame. The zoom factor is computed based on the ratio of the target bounding box width to the current bounding box width:

\[
\text{zoom\_factor} = \frac{\text{target\_width}_{\text{bbox}}}{\text{width}_{\text{bbox}}}
\]

This zoom factor is then used to adjust the camera's zoom level programmatically, ensuring that the eye remains consistently sized within the frame, irrespective of the participant’s distance from the camera. Meanwhile, the iris coordinates, after being scaled back to the original image size, are used to fine-tune the camera’s focal length. The autofocus system dynamically adjusts the focus based on these iris coordinates, ensuring that the camera is precisely focused on the iris region. Once the focus and zoom are set, the eye image is cropped using the eye bounding box coordinates into a 640x480 pixel image, ensuring a consistent and high-quality output suitable for further processing. By implementing a feedback loop that adjusts both the zoom using eye coordinates and the focus using iris coordinates, the application ensures that the eye remains consistently well-framed and sharply focused, resulting in high-quality images. Fig.~\ref{fig:UI_Preview} shows the developed UI and camera preview with detected bounding boxes and the showing real-time feedback to the user on the ongoing process.

After cropping, the images undergo basic image quality assessment checks, as mentioned in Section IV.A2. If these checks are passed, the image frame is then subjected to the ISO/IEC 29794-6 standards for a more rigorous quality evaluation. Should the image meet all the specified quality standards, it is saved to the local storage of the mobile phone. However, if the image fails any of the quality checks, the process loops back to update the camera preview, and the process is restarted again in the loop to capture a new frame by providing real-time feedback to the user. Additionally, we provide an option for segmentation; if the user requests segmentation, the G-ATTU-Net segmentation model is initialized, and the segmentation is performed on the saved image by extracting the respective red channel. The output mask is then saved in the local storage along with the eye image.

\section{Experimental Setup}
\begin{figure}[h]
\centering
    \includegraphics[width=9cm, height=6cm]{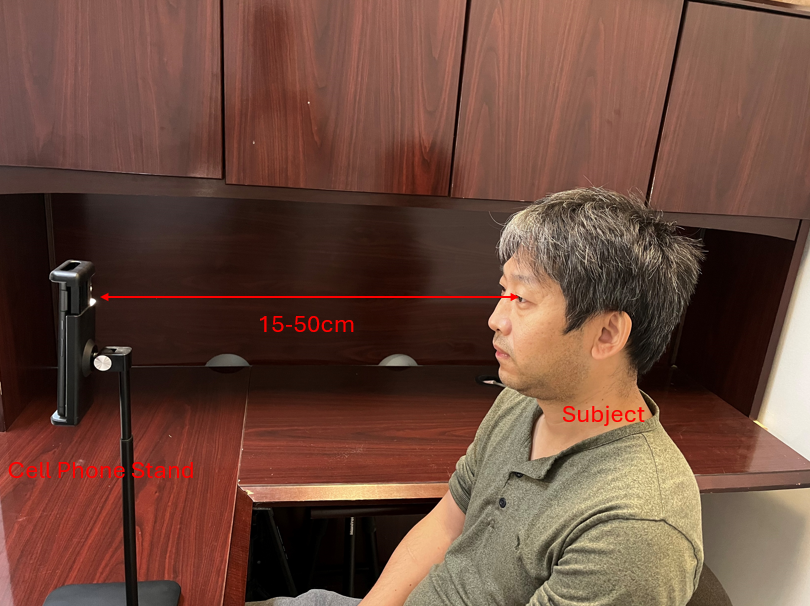}
    \caption{\footnotesize Experimental setup with the mobile device mounted on a stand and the participant seated at varying capture distances.}
    \label{fig:ExperimentalSetup}
    \vspace{-3mm}
 \end{figure}

The experimental setup, illustrated in Fig.~\ref{fig:ExperimentalSetup}, was established for the study. Participants were invited to take part in the experiment, using both the developed application and a commercially available NIR scanner, the IrisGuard IG-AD100 Dual Iris Camera \cite{IrisGuard}. Before participation, each individual underwent a consent process, during which their rights and the data protection policy, ensuring privacy, were thoroughly explained.

The study utilized a Samsung Galaxy S21 Ultra 5G mounted on a mobile stand, positioned as shown in Fig.~\ref{fig:ExperimentalSetup}. The mobile flashlight served as the primary illumination source. Participants were instructed to sit comfortably on a chair, maintaining distances of 25 cm and 50 cm in indoor conditions. The front-facing side of the participants was aligned with a dark wall to minimize reflections and other artifacts introduced by ambient light. Data for each participant were collected in two sessions: the first at a distance of 25 cm and the second at 50 cm. For both sessions, NIR images were also captured using the IrisGuard AD100 scanner.

\begin{figure}[h]
\centering
    \includegraphics[width=9cm, height=6cm]{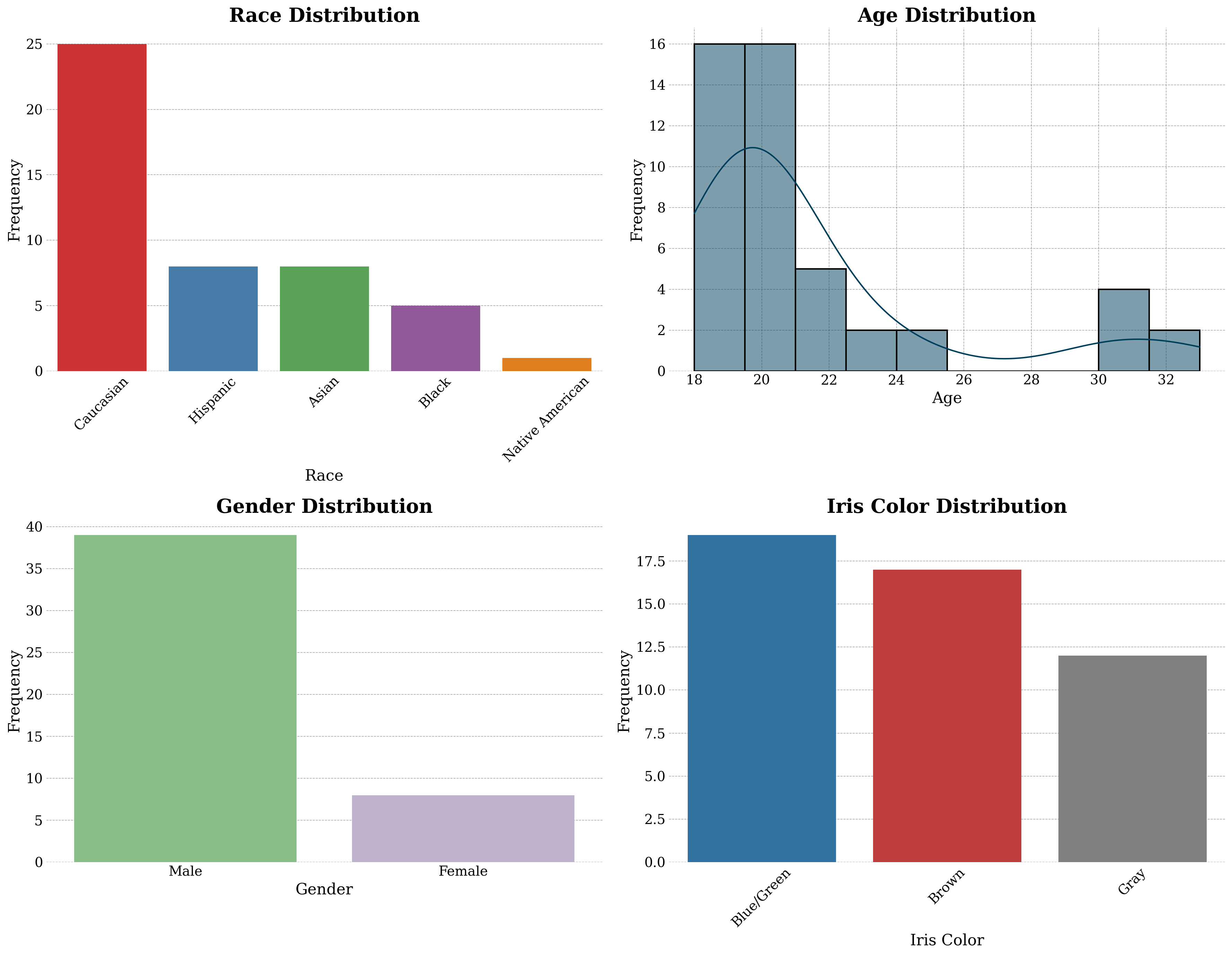}
    \caption{\footnotesize Race, age, gender and iris color distribution in CUVIRIS the database.}
    \label{fig:datasetstats}
    \vspace{-3mm}
 \end{figure}

\section{CUVIRIS Dataset}
\begin{figure*}[h]
\centering
    \includegraphics[width=1\textwidth]{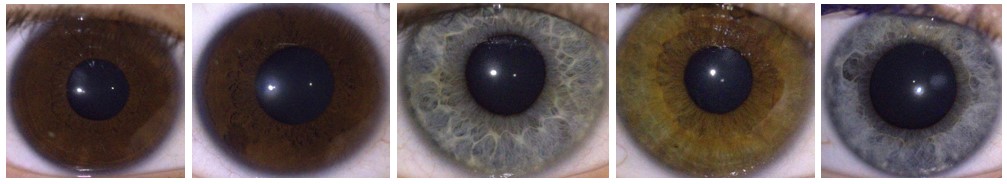}
    \caption{\footnotesize Sample cropped irides images from the CUVIRIS dataset.}
    \label{fig:sample_images}
    \vspace{-3mm}
 \end{figure*}

A total of 47 subjects participated in the study, representing a diverse distribution of race, gender, age, and iris color. Fig.~\ref{fig:datasetstats} illustrates the dataset distribution with respect to these demographic factors. The CUVIRIS dataset contains a total of 188 NIR images from 47 subjects, with 2 images for each left and right eye. The images captured using the Android application include a total of 752 images, with 8 images per eye—4 captured at a distance of 25cm and another 4 at a distance of 50cm. Each participant contributed a total of 20 images: 4 NIR images (2 from each eye) and 16 VIS images (8 from each eye). Of the VIS images, 4 were captured at a distance of 25cm and 4 at 50cm. Fig.~\ref{fig:sample_images} shows sample images captured in both the NIR and visible light spectrum. As shown in the images, high-quality eye captures were consistently achieved, regardless of the capture distance (25 or 50 cm). This dataset will be publicly available for research purposes.

\begin{figure*}[h]
\centering
    \includegraphics[width=1\textwidth]{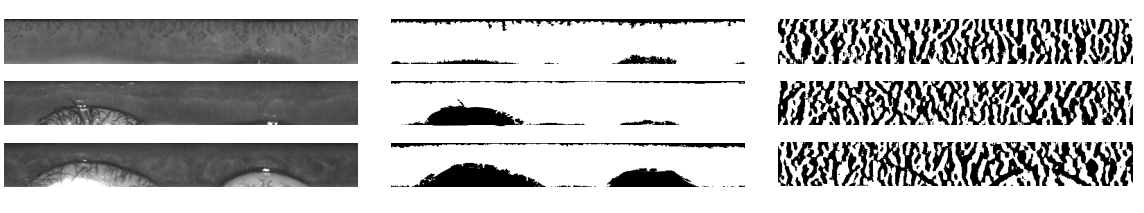}
    \caption{\footnotesize TNormalization and code generation. The left column shows normalized iris images after rubbersheet transformation, the middle column displays the corresponding normalized masks, and the right column presents the iris codes generated by the OSIRIS v4.1 toolkit.}
    \label{fig:iris_code}
    \vspace{-3mm}
 \end{figure*}

\section{Approach}
\subsection{Parameterization and Iris code creation}
The application outputs both the original eye image and the corresponding mask generated by the proposed G-ATTU-Net segmentation model, where on-pixels correspond to the iris and off-pixels to non-iris regions. To generate the normalized iris and mask, we employed the parameterization technique described by Hofbauer et al. \cite{hofbauer2019exploiting}. The process began with detecting the pupillary and limbic boundaries of the iris using the Circular Hough Transform, which accurately identifies these boundaries and represents them as concentric circles. These circles are crucial for performing the rubber sheet transformation, which we applied to normalize the iris by converting its texture from a polar coordinate system to a Cartesian one. This transformation produced a rectangular image of the iris, where each pixel in the normalized image corresponds to a specific region of the original iris. Both the normalized iris and mask were standardized to a fixed size of 512x64 pixels. We ensured that the same transformation was applied to the corresponding mask, allowing us to focus only on the valid regions of the iris—those free from occlusions or noise—in subsequent processing steps.

After normalization, we used the OSIRIS v4.1 toolkit \cite{othman2016osiris} to generate the iris code from the normalized iris image and its corresponding mask. This involved applying a Gabor filter to the normalized iris. Gabor filters are particularly effective for capturing the texture of the iris because they allow analysis at various scales and orientations, extracting fine details characteristic of the iris pattern. The output of the filter was phase information, which we encoded into a binary format, resulting in the iris code. The binary iris code, like the normalized iris and mask, was produced in a 512x64 pixel format, where each bit in the code represents specific texture information derived from the Gabor filter. In OSIRIS v4.1, both the real and imaginary parts of the Gabor filter output were encoded separately, creating a highly detailed and discriminative iris code. Fig.~\ref{fig:iris_code} illustrates this process, displaying the normalized iris (left), the corresponding mask generated using the parameterization technique (middle), and the resulting iris code produced by the OSIRIS toolkit (right).

\subsection{Matching process}
In the iris recognition process, the Hamming Distance (HD) equation, as shown in Equation \ref{eq:hd}, serves as a critical metric for measuring the dissimilarity between two irises. This equation evaluates the variance between binary strings derived from the iris codes and their corresponding masks, as initially outlined by Daugman \cite{daugman2009iris}. By incorporating a mask in the HD calculation, we can achieve a more precise comparison of iris patterns by filtering out noise and artifacts, thereby focusing on the relevant features of the iris. To adhere to standardized procedures, we utilized iris codes and masks processed through the OSIRIS v4.1 toolkit and parameterization technique, respectively.

To generate the final enrollment and verification iris codes, we employed logical operations as defined in Equations \ref{eq:enrollment} and \ref{eq:verification}. Specifically, we applied a logical AND operation to merge each iris code with its corresponding mask, effectively reducing the impact of variations or distortions in the iris images caused by factors such as lighting conditions or occlusions. In these equations, `iriscode1' and `mask1' represent the enrollment iris code and mask, while `iriscode2' and `mask2' represent the verification iris code and mask. The Hamming Distance, calculated as shown in Equation \ref{eq:hd}, determines the degree of dissimilarity between the two irises by comparing these processed codes.

\begin{equation}
HD = \frac{\lVert(\text{iriscode1} \oplus \text{iriscode2}) \cap \text{mask1} \cap \text{mask2}\rVert}{\lVert\text{mask1} \cap \text{mask2}\rVert}
\label{eq:hd}
\end{equation}

\begin{equation}
\text{enrollment} = \text{iriscode1} \cap \text{mask1}
\label{eq:enrollment}
\end{equation}

\begin{equation}
\text{verification} = \text{iriscode2} \cap \text{mask2}
\label{eq:verification}
\end{equation}

During the matching process, we accounted for potential rotational inconsistencies between the iris templates. To mitigate this issue, one of the templates was shifted both left and right in a bit-wise manner. Specifically, we performed ±7 bit-wise shifts on the gallery dataset to ensure that the probe was matched against all shifted samples. This shifting process helps to align the templates accurately, thus improving the reliability and accuracy of the iris recognition results.

\begin{figure*}[h]
\centering
    \includegraphics[width=\textwidth]{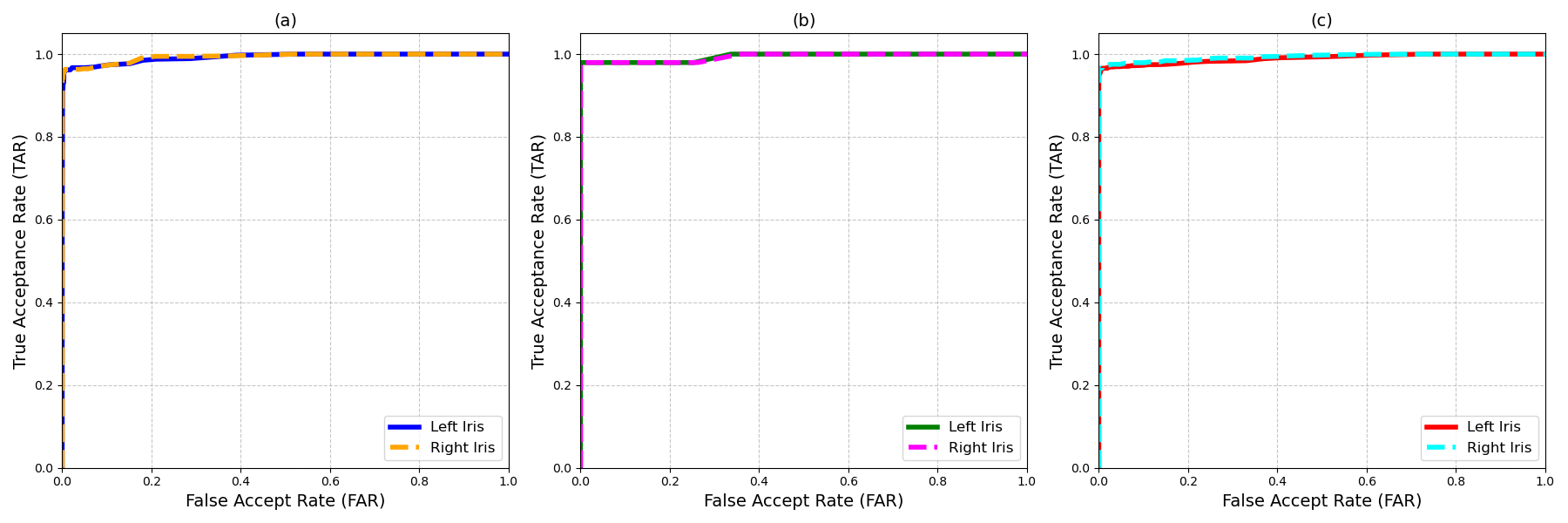}
    \caption{\footnotesize ROC curves for various comparisons: (a) visible light, (b) NIR, and (c) cross-spectral.}
    \label{fig:ROC_all}
    \vspace{-3mm}
 \end{figure*}

\begin{figure}[h]
\centering
    \includegraphics[width=9cm, height=6cm]{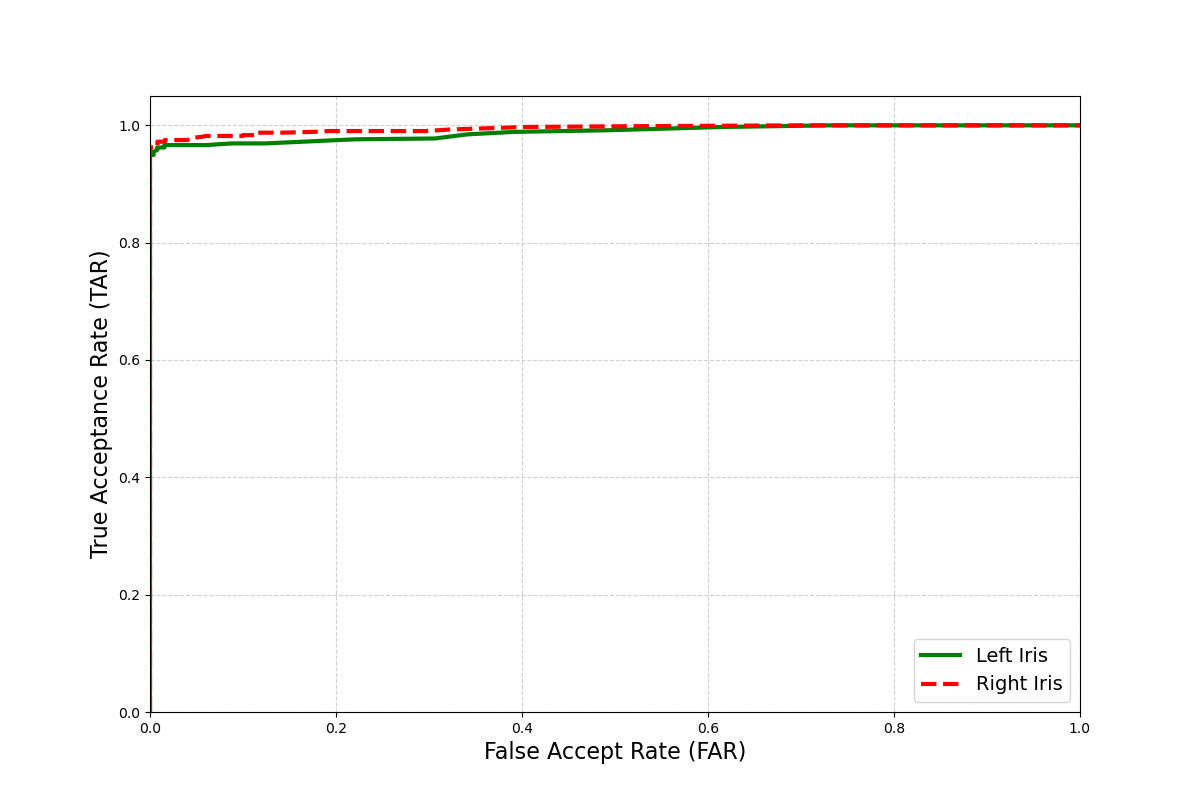}
    \caption{\footnotesize ROC curve for distance-based comparison.}
    \label{fig:ROC_distance}
    \vspace{-3mm}
 \end{figure}

\begin{figure*}[h]
\centering
    \includegraphics[width=\textwidth]{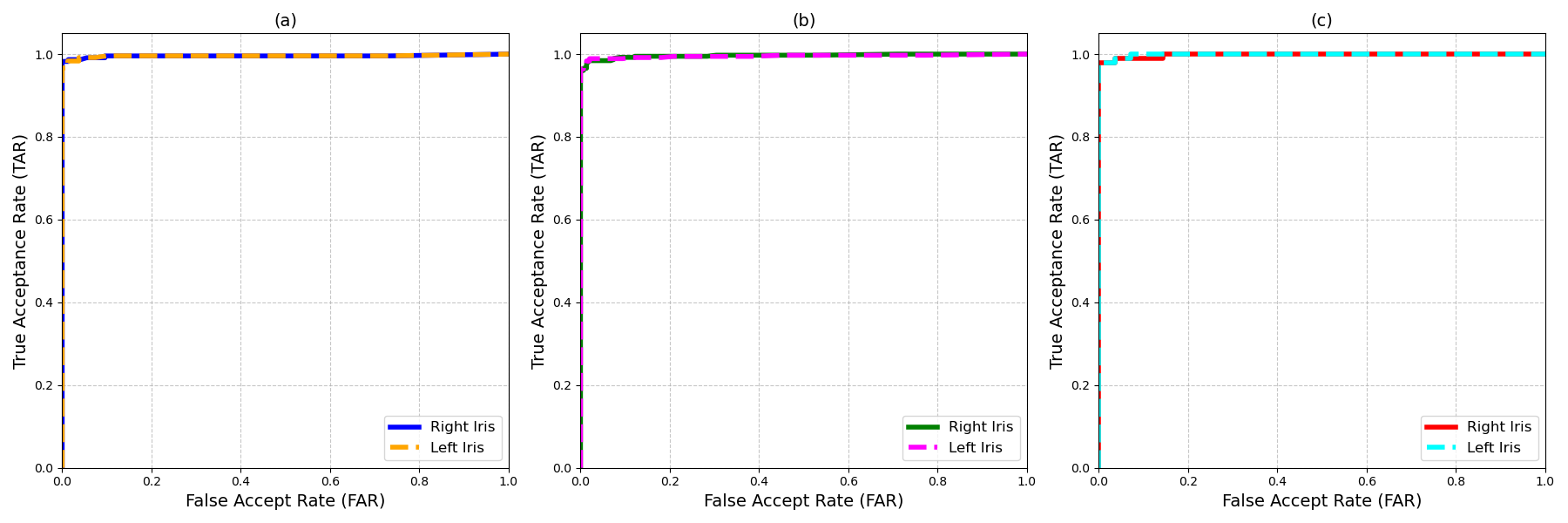}
    \caption{\footnotesize ROC curves for color-based comparison: (a) blue, (b) brown, and (c) gray irises.}
    \label{fig:ROC_color}
    \vspace{-3mm}
 \end{figure*}

\section{Experiments on CUVIRIS Dataset}
To evaluate the performance of the developed application and segmentation techniques, we conducted verification and identification experiments using the CUVIRIS dataset. Performance was assessed in terms of True Acceptance Rate (TAR) with the False Acceptance Rate (FAR) fixed at 0.01\%. As described in Section VI, the CUVIRIS dataset comprises iris images from 47 subjects, captured in both the NIR and VIS. Each subject's data includes 2 NIR images per eye (left and right) and 16 VIS images—8 images per eye, with 4 captured at a distance of 25 cm and 4 at 50 cm.

For genuine comparison, the first image captured was used as the enrollment template, while the remaining images of the corresponding subject were treated as verification templates. For imposter comparison, a randomly selected enrollment template from one subject was compared against all other subjects' images as verification templates.

First, we analyzed the performance of the VIS iris images, achieving a TAR of 96.57\% for the left iris and 96.81\% for the right iris. Subsequently, the NIR images were compared, yielding a TAR of 97.95\% for both the left and right irises. We then performed cross-spectral comparisons, where NIR images were compared with VIS images. In this case, an NIR image served as the enrollment image, while all corresponding VIS iris images of the same subject were used as verification templates for genuine comparison. For imposter comparison, a NIR image of one subject was compared against all other subjects' visible light iris images. The cross-spectral comparison achieved a TAR of 96.17\% for the left iris and 96.31\% for the right iris. Fig.~\ref{fig:ROC_all} presents the ROC curves for these comparisons.

Additionally, we assessed the impact of capture distance on performance by comparing images taken at 25 cm and 50 cm. For genuine comparison, one randomly selected sample captured at 25 cm was used as the enrollment image, and all remaining images taken at 50 cm were used as verification images. For imposter comparison, one random sample from each subject captured at 25 cm was compared against all other subjects' images captured at 50 cm, excluding the enrollment image. The distance-based comparison resulted in a TAR of 96.22\% for the left iris and 97.21\% for the right iris. Fig.~\ref{fig:ROC_distance} presents the ROC curves obtained for distance-based comparisons.

Finally, we analyzed the performance based on iris color. The CUVIRIS dataset includes 15 blue, 26 brown, and 6 gray-colored irises. The TAR for blue irises was 96.67\% for the left iris and 98.23\% for the right iris. For brown irises, the TAR was 96.41\% for the left iris and 96.05\% for the right iris. Similarly, for gray irises, the TAR was 97.92\% for both the left and right irises. Fig.~\ref{fig:ROC_color} presents the DET plots for various iris colors. Overall, the consistent TAR results across different comparisons demonstrate the robustness and effectiveness of our proposed method.

\section{Discussion}
The experiments conducted on the CUVIRIS dataset have demonstrated the effectiveness of the proposed Android application and segmentation techniques in achieving high performance in iris recognition tasks under varying conditions. The True Acceptance Rate (TAR) for VIS images was 96.57\% for the left iris and 96.81\% for the right iris, while the TAR for NIR images was 97.95\% for both eyes. These results indicate that the proposed system performs reliably across different spectra.

Compared to previous studies that utilized VIS iris images, such as the work by Raja et al. [12], which reported a Genuine Match Rate (GMR) of 91.01\% at a False Match Rate (FMR) of 0.01\%, our approach shows significant improvements in accuracy. Additionally, the segmentation accuracy achieved by our G-ATTU-Net model is competitive with existing deep learning-based segmentation methods while offering the advantage of a smaller model size, making it feasible for deployment on resource-constrained mobile devices.

The cross-spectral comparisons, where NIR images were compared with VIS images, achieved a TAR of 96.17\% for the left iris and 96.31\% for the right iris. This consistency across different spectra highlights the robustness of the proposed method. Furthermore, the distance-based analysis, comparing images captured at 25 cm and 50 cm, yielded TARs of 96.22\% and 97.21\% for the left and right irises, respectively. This demonstrates that the system maintains high accuracy regardless of the capture distance, which is critical for practical applications where user positioning may vary. Our analysis also revealed that iris color did not significantly impact the recognition accuracy, with TARs consistently above 96\% across blue, brown, and gray irises. This suggests that the proposed system is well-suited for a diverse user base, regardless of iris pigmentation.

The high accuracy rates achieved across various conditions suggest that the proposed method is highly applicable in real-world scenarios, particularly for enhancing smartphone security. The lightweight nature of the G-ATTU-Net model further reinforces its suitability for mobile platforms, offering a balance between computational efficiency and segmentation accuracy.

Despite these promising results, certain limitations should be acknowledged. For instance, the study focused on a relatively small and controlled dataset, which may not fully capture the variability encountered in broader real-world applications. Additionally, while the system performed well in indoor conditions, further testing in outdoor or more challenging lighting environments would be necessary to confirm its robustness in those scenarios. Future research could explore expanding the dataset to include a more diverse range of subjects and environmental conditions. Additionally, optimizing the G-ATTU-Net model for even greater efficiency, possibly through the use of novel quantization techniques, could further enhance its deployment potential in low-power devices.

\section{Conclusion}

In this study, we developed and evaluated a high-quality visible light iris recognition system for smartphones, incorporating a lightweight segmentation model, G-ATTU-Net, optimized for resource-constrained environments. The system was rigorously tested on the CUVIRIS dataset, demonstrating robust performance across various conditions, including different spectra, distances, and iris colors. The results show that the proposed method achieves high True Acceptance Rates (TAR) across all tested conditions, with particularly strong performance in both visible and NIR spectrums. The consistent accuracy observed in cross-spectral and distance-based comparisons underscores the method's versatility and potential applicability in real-world scenarios, such as mobile device security. While the study was conducted on a controlled dataset, the promising results suggest that the proposed approach is a viable solution for real-time iris recognition on mobile platforms. Future work will focus on expanding the dataset to include a broader range of environmental conditions and further optimizing the model for even greater efficiency. In conclusion, the proposed system represents a significant step forward in mobile iris recognition technology, offering a practical and effective solution for enhancing biometric security on smartphones.
 
\bibliographystyle{IEEEtran}\
\bibliography{bibliography}

\begin{thebibliography}{10}
\providecommand{\url}[1]{#1}
\csname url@samestyle\endcsname
\providecommand{\newblock}{\relax}
\providecommand{\bibinfo}[2]{#2}
\providecommand{\BIBentrySTDinterwordspacing}{\spaceskip=0pt\relax}
\providecommand{\BIBentryALTinterwordstretchfactor}{4}
\providecommand{\BIBentryALTinterwordspacing}{\spaceskip=\fontdimen2\font plus
\BIBentryALTinterwordstretchfactor\fontdimen3\font minus \fontdimen4\font\relax}
\providecommand{\BIBforeignlanguage}[2]{{%
\expandafter\ifx\csname l@#1\endcsname\relax
\typeout{** WARNING: IEEEtran.bst: No hyphenation pattern has been}%
\typeout{** loaded for the language `#1'. Using the pattern for}%
\typeout{** the default language instead.}%
\else
\language=\csname l@#1\endcsname
\fi
#2}}
\providecommand{\BIBdecl}{\relax}
\BIBdecl

\bibitem{markert2020pin}
P.~Markert, D.~V. Bailey, M.~Golla, M.~D{\"u}rmuth, and A.~J. Aviv, ``This pin can be easily guessed: Analyzing the security of smartphone unlock pins,'' in \emph{2020 IEEE Symposium on Security and Privacy (SP)}.\hskip 1em plus 0.5em minus 0.4em\relax IEEE, 2020, pp. 286--303.

\bibitem{spolaor2016biometric}
R.~Spolaor, Q.~Li, M.~Monaro, M.~Conti, L.~Gamberini, and G.~Sartori, ``Biometric authentication methods on smartphones: A survey.'' \emph{PsychNology Journal}, no.~2, 2016.

\bibitem{bowyer2008image}
K.~W. Bowyer, K.~Hollingsworth, and P.~J. Flynn, ``Image understanding for iris biometrics: A survey,'' \emph{Computer vision and image understanding}, vol. 110, no.~2, pp. 281--307, 2008.

\bibitem{daugman2009iris}
J.~Daugman, ``How iris recognition works,'' in \emph{The essential guide to image processing}.\hskip 1em plus 0.5em minus 0.4em\relax Elsevier, 2009, pp. 715--739.

\bibitem{raja2015smartphone}
K.~B. Raja, R.~Raghavendra, V.~K. Vemuri, and C.~Busch, ``Smartphone based visible iris recognition using deep sparse filtering,'' \emph{Pattern Recognition Letters}, vol.~57, pp. 33--42, 2015.

\bibitem{hosseini2010pigment}
M.~S. Hosseini, B.~N. Araabi, and H.~Soltanian-Zadeh, ``Pigment melanin: Pattern for iris recognition,'' \emph{IEEE transactions on instrumentation and measurement}, vol.~59, no.~4, pp. 792--804, 2010.

\bibitem{winston2019comprehensive}
J.~J. Winston and D.~J. Hemanth, ``A comprehensive review on iris image-based biometric system,'' \emph{Soft Computing}, vol.~23, pp. 9361--9384, 2019.

\bibitem{rattani2016icip}
A.~Rattani, R.~Derakhshani, S.~K. Saripalle, and V.~Gottemukkula, ``Icip 2016 competition on mobile ocular biometric recognition,'' in \emph{2016 IEEE international conference on image processing (ICIP)}.\hskip 1em plus 0.5em minus 0.4em\relax IEEE, 2016, pp. 320--324.

\bibitem{de2015mobile}
M.~De~Marsico, M.~Nappi, D.~Riccio, and H.~Wechsler, ``Mobile iris challenge evaluation (miche)-i, biometric iris dataset and protocols,'' \emph{Pattern Recognition Letters}, vol.~57, pp. 17--23, 2015.

\bibitem{santos2015fusing}
G.~Santos, E.~Grancho, M.~V. Bernardo, and P.~T. Fiadeiro, ``Fusing iris and periocular information for cross-sensor recognition,'' \emph{Pattern Recognition Letters}, vol.~57, pp. 52--59, 2015.

\bibitem{edwards2012quantitative}
M.~Edwards, A.~Gozdzik, K.~Ross, J.~Miles, and E.~J. Parra, ``Quantitative measures of iris color using high resolution photographs,'' \emph{American journal of physical anthropology}, vol. 147, no.~1, pp. 141--149, 2012.

\bibitem{proencca2009ubiris}
H.~Proen{\c{c}}a, S.~Filipe, R.~Santos, J.~Oliveira, and L.~A. Alexandre, ``The ubiris. v2: A database of visible wavelength iris images captured on-the-move and at-a-distance,'' \emph{IEEE Transactions on Pattern Analysis and Machine Intelligence}, vol.~32, no.~8, pp. 1529--1535, 2009.

\bibitem{iris_standard_report}
``{ISO/IEC 29794-6:2015 Information Technology - Biometric Sample Quality -Part 6: Iris Image Data},'' ISO(International Organization for Standardization) and IEC (International Electrotechnical Commission.

\bibitem{IrisGuard}
``{IrisGuard},'' \url{https://www.irisguard.com}, accessed: 2019-11-147.

\bibitem{raja2015iris}
K.~B. Raja, R.~Raghavendra, and C.~Busch, ``Iris imaging in visible spectrum using white led,'' in \emph{2015 IEEE 7th International Conference on Biometrics Theory, Applications and Systems (BTAS)}.\hskip 1em plus 0.5em minus 0.4em\relax IEEE, 2015, pp. 1--8.

\bibitem{proenca2009iris}
H.~Proenca, ``Iris recognition: On the segmentation of degraded images acquired in the visible wavelength,'' \emph{IEEE Transactions on Pattern Analysis and Machine Intelligence}, vol.~32, no.~8, pp. 1502--1516, 2009.

\bibitem{trokielewicz2016iris}
M.~Trokielewicz, ``Iris recognition with a database of iris images obtained in visible light using smartphone camera,'' in \emph{2016 IEEE International Conference on Identity, Security and Behavior Analysis (ISBA)}.\hskip 1em plus 0.5em minus 0.4em\relax IEEE, 2016, pp. 1--6.

\bibitem{adarsh2020yolo}
P.~Adarsh, P.~Rathi, and M.~Kumar, ``Yolo v3-tiny: Object detection and recognition using one stage improved model,'' in \emph{2020 6th international conference on advanced computing and communication systems (ICACCS)}.\hskip 1em plus 0.5em minus 0.4em\relax IEEE, 2020, pp. 687--694.

\bibitem{folk2011overview}
M.~Folk, G.~Heber, Q.~Koziol, E.~Pourmal, and D.~Robinson, ``An overview of the hdf5 technology suite and its applications,'' in \emph{Proceedings of the EDBT/ICDT 2011 workshop on array databases}, 2011, pp. 36--47.

\bibitem{arsalan2017deep}
M.~Arsalan, H.~G. Hong, R.~A. Naqvi, M.~B. Lee, M.~C. Kim, D.~S. Kim, C.~S. Kim, and K.~R. Park, ``Deep learning-based iris segmentation for iris recognition in visible light environment,'' \emph{Symmetry}, vol.~9, no.~11, p. 263, 2017.

\bibitem{jan2024iris}
F.~Jan, S.~Alrashed, and N.~Min-Allah, ``Iris segmentation for non-ideal iris biometric systems,'' \emph{Multimedia Tools and Applications}, vol.~83, no.~5, pp. 15\,223--15\,251, 2024.

\bibitem{zhang2019robust}
W.~Zhang, X.~Lu, Y.~Gu, Y.~Liu, X.~Meng, and J.~Li, ``A robust iris segmentation scheme based on improved u-net,'' \emph{IEEE access}, vol.~7, pp. 85\,082--85\,089, 2019.

\bibitem{lian2018attention}
S.~Lian, Z.~Luo, Z.~Zhong, X.~Lin, S.~Su, and S.~Li, ``Attention guided u-net for accurate iris segmentation,'' \emph{Journal of Visual Communication and Image Representation}, vol.~56, pp. 296--304, 2018.

\bibitem{hofbauer2019exploiting}
H.~Hofbauer, E.~Jalilian, and A.~Uhl, ``Exploiting superior cnn-based iris segmentation for better recognition accuracy,'' \emph{Pattern Recognition Letters}, vol. 120, pp. 17--23, 2019.

\bibitem{othman2016osiris}
N.~Othman, B.~Dorizzi, and S.~Garcia-Salicetti, ``Osiris: An open source iris recognition software,'' \emph{Pattern recognition letters}, vol.~82, pp. 124--131, 2016.

\end{thebibliography}

\end{document}